\documentclass[pra,groupedaddress,superscriptaddress,twocolumn,amsmath,amssymb,a4paper]{revtex4}
\usepackage{geometry}
\usepackage[dvipsnames]{xcolor}
\usepackage{graphicx}
\usepackage{verbatim}
\usepackage{float}
\usepackage{graphics}
\usepackage{mathrsfs}
\usepackage{amssymb}
\usepackage{amsmath}
\usepackage{amsthm}
\usepackage{bm}
\usepackage{dsfont}
\usepackage{hyperref}
\usepackage{braket}
\usepackage{wrapfig}
\usepackage{amsbsy}

\usepackage[T1]{fontenc}

\newcommand{\ave}[1]{\langle #1 \rangle}
\newcommand{\ip}[2]{{\langle #1|}{ #2 \rangle }}

\newcommand{\be}{\begin{eqnarray}}
\newcommand{\ee}{\end{eqnarray}}

\begin{document}

\title{Quantum master equations for a system interacting with quantum gas in the low density limit and for the semiclassical collision model}

\author{S. N. Filippov}
\affiliation{Department of Mathematical Methods for Quantum
Technologies, Steklov Mathematical Institute of Russian Academy of
Sciences, Gubkina St. 8, Moscow 119991, Russia}
\affiliation{Valiev Institute of Physics and Technology of Russian
Academy of Sciences, Nakhimovskii Pr. 34, Moscow 117218, Russia}
\affiliation{Moscow Institute of Physics and Technology,
Institutskii Per. 9, Dolgoprudny, Moscow Region 141700, Russia}

\author{G. N. Semin}
\affiliation{Moscow Institute of Physics and Technology,
Institutskii Per. 9, Dolgoprudny, Moscow Region 141700, Russia}

\author{A. N. Pechen}
\affiliation{Department of Mathematical Methods for Quantum Technologies, Steklov Mathematical Institute of Russian Academy of
Sciences, Gubkina St. 8, Moscow 119991, Russia}
\affiliation{National University of Science and Technology
``MISIS'', Leninskii Pr. 4, Moscow, 119049, Russia}

\begin{abstract}
A quantum system interacting with a dilute gas experiences
irreversible dynamics. The corresponding master equation can be
derived within two different approaches: The fully quantum
description in the low-density limit and the semiclassical
collision model, where the motion of gas particles is classical
whereas their internal degrees of freedom are quantum. The two
approaches have been extensively studied in the literature, but
their predictions have not been compared. This is mainly due to
the fact that the low-density limit is extensively studied for
mathematical physics purposes, whereas the collision models have
been essentially developed for quantum information tasks such as a
tractable description of the open quantum dynamics. Here we
develop and for the first time compare both approaches for a spin
system interacting with a gas of spin particles. Using some
approximations, we explicitly find the corresponding master
equations including the Lamb shifts and the dissipators. The low
density limit in the Born approximation for fast particles is
shown to be equivalent to the semiclassical collision model in the
stroboscopic approximation. We reveal that both approaches give
exactly the same master equation if the gas temperature is high
enough. This allows to interchangeably use complicated
calculations in the low density limit and rather simple
calculations in the collision model.
\end{abstract}

\maketitle

\section{Introduction}
Any realistic quantum system is open because of unavoidable
coupling to its environment. The theory of open quantum systems
studies the effect of the surrounding environment on the system
dynamics~\cite{breuer-2002}. The environment can be represented as
a large reservoir either in thermodynamic
equilibrium~\cite{schoeller-2018} or in a non-equilibrium state.
The system-reservoir interaction entangles the system with the
environmental degrees of freedom, which typically leads to the
irreversible system decoherence. Such a decoherence significantly
affects quantum transport~\cite{cui-2006,talarico-2019}, molecular
excitation dynamics and relaxation~\cite{valkunas-2013}, and
performance of quantum sensors~\cite{degen-2017}. It is the
decoherence that complicates the protocols of quantum information
transmission~\cite{wilde-2017} and
processing~\cite{nielsen-2000,filippov-2019}. This circumstance
makes the study of decoherence an important field of research for
the development of quantum technologies~\cite{glaser-2015}.

The state of a quantum system is represented by the density
operator $\varrho(t)$ that is a Hermitian positive-semidefinite
unit-trace operator acting in the system Hilbert space ${\cal H}$.
Let ${\cal T}({\cal H})$ be the space of trace class operators
acting in ${\cal H}$. The open dynamics is usually described by
the time-convolutionless master equation $\frac{d}{dt}\varrho(t) =
{\cal L}_t [\varrho(t)]$, which is obtained by averaging over the
environmental degrees of freedom in the joint evolution of the
system and the reservoir. The generator ${\cal L}_t: {\cal
T}({\cal H}) \mapsto {\cal T}({\cal H})$ is time-dependent in
general, which may lead to non-Markovian
effects~\cite{rivas-2014,breuer-2016,de-vega-2017,benatti-2017,fc-2018,li-2018,luchnikov-2019}.
There are physical situations, however, where the generator is
time-independent within the characteristic timescale of system
evolution. Microscopic derivations of the master equation
\begin{equation} \label{master-equation}
\frac{d}{dt}\varrho(t) = {\cal L} [\varrho(t)]
\end{equation}
can be obtained in the weak coupling
limit~\cite{van-hove-1954,davies-1974,spohn-1978,accardi-1990},
the singular coupling limit~\cite{palmer-1977,gorini-1978}, the
stochastic limit~\cite{accardi-book,pechen-2002}, the low density
limit for gas
environment~\cite{dumcke-1985,accardi-1991,accardi-1992,rudnicki-1992,apv-2002,accardi-2003,pechen-2004,pechen-jmp-2006},
the stroboscopic limit in the collision
model~\cite{rau-1963,giovannetti-2012,lorenzo-2017,luchnikov-2017},
and the monitoring approach to derivation of linear Boltzmann
equation~\cite{hornberger-2007,hornberger-2008,vacchini-2009,smirne-2010}.
In all these approximations, the particular form of ${\cal L}$ is
expressed through the system-environment interaction Hamiltonian
and the reservoir equilibrium state. The solution of the master
equation \eqref{master-equation} is given by the quantum dynamical
semigroup $e^{{\cal L}t}$, whose complete positivity makes the
generator ${\cal L}$ take the
Gorini-Kossakowski-Sudarshan-Lindblad (GKSL)
form~\cite{gks-1976,lindblad-1976}:
\begin{equation} \label{GKSL}
{\cal L} [\varrho] = - \frac{i}{\hbar} [H,\varrho] + \sum_{k}
\gamma_k \left( A_k \varrho A_k^{\dag} - \frac{1}{2}
\{A_k^{\dag}A_k,\varrho\} \right),
\end{equation}
Here $[\cdot,\cdot]$ and $\{\cdot,\cdot\}$ denote the
commutator and anticommutator, respectivey, $H$ is a Hermitian
operator, $\gamma_k > 0$ is the relaxation rate for the
$k$th channel of decoherence, and $\{A_k\}$ are the jump
operators.

In this paper, we consider a quantum system interacting with a gas
reservoir. The gas is supposed to be dilute, so that gas particles
rarely interact with the system. The scattering of gas particles
on the system leads to the system decoherence. Such a situation
takes place in all vacuum experiments because of the presence of a
background gas, e.g., in levitated
optomechanics~\cite{hornberger-2008,martinetz-2018}, ion
traps~\cite{wineland-1998,serra-2001}, and atom
interferometers~\cite{uys-2005}. Finding the specific form of the
generator ${\cal L}$ and determining the relaxation rates is an
important timely problem for control and
manipulation~\cite{pechen-2006,pechen-rabitz-2014,pechen-2019} of
quantum systems in the presence of a background gas.

There are two distinctive theoretical approaches to treat motional
degrees of freedom for gas particles: (i) quantum and (ii)
classical.

Within the first approach, the reservoir is treated as an ensemble
of non-interacting quantum particles being in the Gibbs state
$\rho_{\rm R}=Z^{-1}\exp[{-\beta(H_{\rm R}-\mu \hat{N})}]$ with
inverse temperature $\beta$ and chemical potential $\mu$, where
$Z$ is the normalizing factor, $\hat{N}$ is the number operator
for gas particles, and $H_{\rm R}$ is the free gas Hamiltonian.
The reservoir can be in a non-equilibrium Gaussian state in
general. The interaction between the system and gas particles has
the scattering type and preserves the number of gas particles,
i.e., commutes with $\hat{N}$. Due to interaction with the system,
gas particles are scattered on the system and this scattering
induces transitions between the system's quantum states. The basic
assumption for the ab-initio derivation of the master
equation~\eqref{master-equation} within   this approach is that
density of gas particles $n$ is low so that only collisions
between the system and one particle of the gas dominate. The
interaction of the system simultaneously with two or more gas
particles is assumed to have negligible probability. Formally,
this assumption is described by taking the limit $n \to + 0$.
However, simply taking this limit would imply complete
disregarding of the reservoir and lead to a trivial system
dynamics. To get a non-trivial dynamics, one has to also consider
long time scale $t \approx 1/n \to +\infty$. Thus  the \emph{low
density limit} (LDL) is defined as the following joint limit: the
gas density $n \rightarrow +0$, the time $t \rightarrow +\infty$,
such that $nt$ is fixed (it is the new slow time scale). The
explicit form of the generator~\eqref{GKSL} in the LDL is derived
{\it ab initio} from exact microscopic dynamics without any
further assumptions and is expressed through the scattering
$T$-matrix for interaction of the system and one gas particle in
Refs.~\cite{dumcke-1985,accardi-1991,apv-2002,accardi-2003,pechen-2004}
and is briefly reviewed in Ref.~\cite{breuer-2002}, Sec. 3.3.4.
The approach of the authors of
Refs.~\cite{apv-2002,accardi-2003,pechen-2004,pechen-jmp-2006}
allows to derive not only the master equation for the reduced
dynamics, but a full quantum stochastic differential equation for
the approximate unitary dynamics of the system and quantum gas.
Important is that the interaction between the system and the gas
is generally considered to be strong and fully quantum mechanical.
Thus, the LDL allows to derive a tractable master equation for a
fully quantum system in the strong coupling regime (beyond the
perturbation expansion).

Within the second approach, the gas particles move along the
classical trajectories whereas their internal degrees of freedom
are
quantum~\cite{alicki-2003,koniorczyk-2008,vacchini-2009,smirne-2010}
(similarly to the micromaser theory~\cite{rempe-1990}). As a
result, the interaction between the quantum system and the
reservoir particle is only activated during the collision time $\tau$;
the system-particle interaction energy increases up to the
characteristic value $U_0$ during the collision (when the system
and the particle are close to each other) and vanishes prior and
after the collision (when the system and the particle are far
apart). Since the reservoir is large and the gas is dilute, each
gas particle interacts with the system at most once and one can
neglect simultaneous collisions of the system with several
particles. This feature is similar to the LDL approach. The master
equation~\eqref{master-equation} was obtained for such a
semiclassical \emph{collision model} (CM) in the stroboscopic
approximation $U_0 \tau \ll \hbar$ (see, e.g.,
Refs.~\cite{rau-1963,giovannetti-2012,lorenzo-2017,scarani-2002,rybar-2012,mccloskey-2014,kretschmer-2016,dabrowska-2017,filippov-2017,ciccarello-2017},
where the generator ${\cal L}_t$ is derived for rectangular
activation functions, various interaction types, and environment
states).

Interestingly, the predictions of these two approaches have not
been compared in the literature. This is mainly due to the fact
that the LDL approach was extensively studied in mathematical
physics, whereas the collision models have been essentially
developed for quantum information tasks as a tractable description
of the open quantum dynamics. However, the common dominating role
of the simultaneous interaction of the system with at most one gas
particle and the absence of many-body interactions makes such a
comparison a natural task. The goal of this paper is to fill the
gap between the two approaches and provide the conditions under
which these approaches lead to the same resulting master equation.
We consider the system and gas particles as having internal
degrees of freedom and establish equivalence, under certain
conditions, between the master equations derived using LDL and CM.
It is worth mentioning that a master equation describing
collisional decoherence for systems with internal degrees of
freedom was derived also using a scattering description of the
interaction events~\cite{hornberger-2007,smirne-2010}. The
established in our work equivalence relation simplifies the
analysis of such open quantum systems for which either of the
models is easy to handle. For instance, one can use the
stroboscopic approximation in the collision model for fast
particles in some thermodynamic
problems~\cite{levy-2012,kosloff-2013,kosloff-2019} instead of
dealing with the fully quantum description.

\begin{figure}
\centering
\includegraphics[width=8cm]{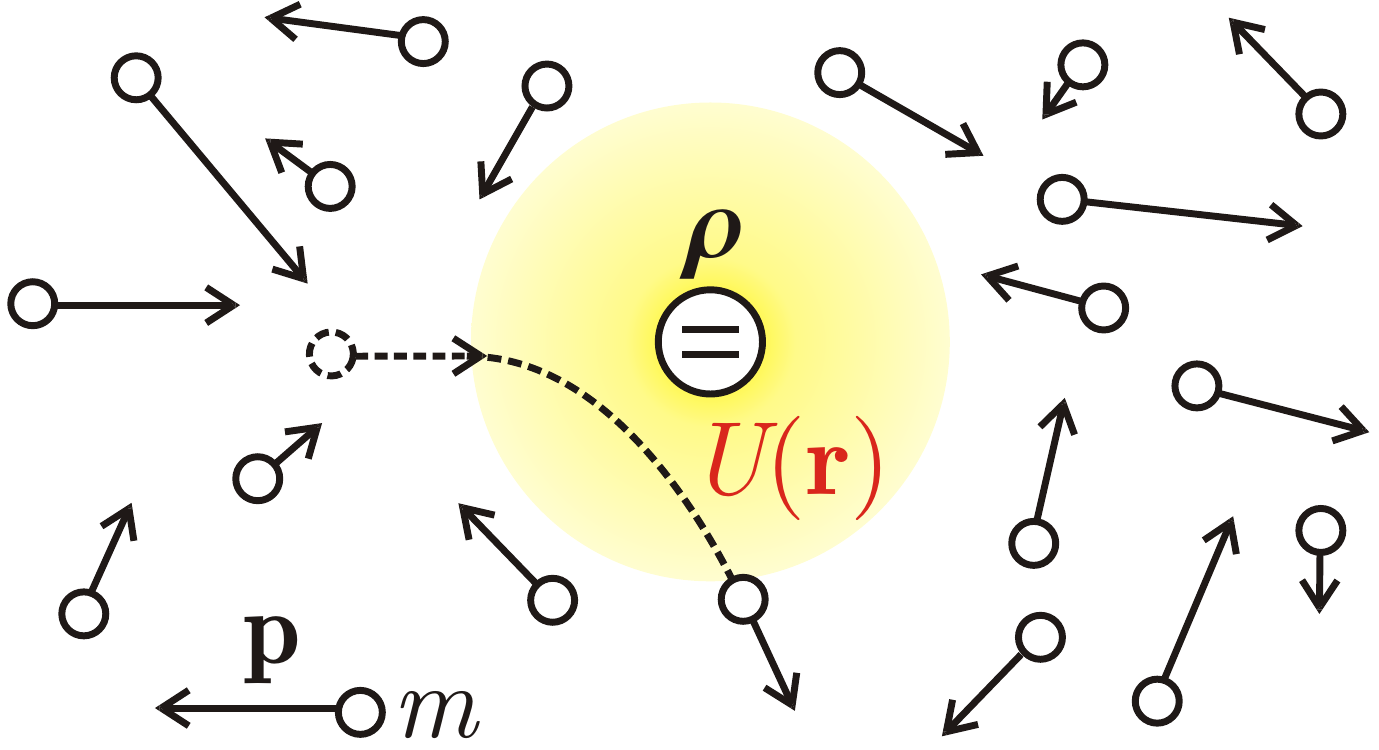}
\caption{Open dynamics of the system (large circle) with density
operator $\varrho$ due to interaction with a diluted gas (small
circles).} \label{figure1}
\end{figure}

To take into account only the relevant physical parameters, we
consider a simplified model of elastic collisions and an
energy-degenerate quantum system. This model describes, for
instance, a quantum spin system interacting via collisions with
spin gas particles (see Fig.~\ref{figure1}).

The paper is organized as follows. In Sec.~\ref{section-ldl}, we
review the LDL model and derive the explicit form of the generator
${\cal L}^{\rm LDL}$ for the case when gas particles have internal
degrees of freedom. In Sec.~\ref{section-cm}, we review the
collision models with a factorized environment and derive the
generator ${\cal L}^{\rm CM}$ for the case of fast particles, when
the trajectories of gas particles can be considered as straight
lines. In Sec.~\ref{section-comparison}, we compare the results of
Secs.~\ref{section-ldl} and~\ref{section-cm} and find the
conditions for their equivalence. In
Sec.~\ref{section-conclusions}, conclusions are given.

\section{The low density limit for the fully quantum model}
\label{section-ldl}

\subsection{Gas of particles with no internal degrees of freedom}

Consider an ideal gas of $N$ nonrelativistic particles each of
mass $m$ moving in $\mathbb{R}^3$. Thermal state of such gas is
described by the density operator
\begin{equation} \label{rho-env}
\varrho_{E} = \varrho_1^{\otimes N}, \quad \varrho_1 =
\frac{(2\pi\hbar)^3}{V} \int f({\bf p}) \ket{\bf p} \bra{\bf p}
d^3{\bf p},
\end{equation}
where $V$ is a volume occupied by gas, $\ket{{\bf p}}$ is a
single-particle state with the definite momentum ${\bf p}$ such
that $\ip{{\bf p}}{{\bf p}'} = \delta({\bf p} - {\bf p}')$, and
$f({\bf p})$ is the Maxwell--Boltzmann distribution
\begin{equation} \label{p-distribution}
f({\bf p}) = \left( 2 \pi m k T \right)^{-3/2} \exp \left(
-\frac{{\bf p}^2}{2mkT} \right).
\end{equation}
Here $k$ is the Boltzmann constant and $T$ is the
temperature. In the position representation, we have
\begin{equation}
\ip{\bf r}{\bf p} = (2\pi\hbar)^{-3/2} \exp \left( \frac{i{\bf
p}{\bf r}}{\hbar} \right),
\end{equation}
so the density operator~\eqref{rho-env} is properly
normalized, namely,
\begin{equation}
{\rm tr}[\varrho_{E}] = \left( \int \bra{\bf r} \varrho_{E}
\ket{\bf r} d^3{\bf r} \right)^N = \left( \int f({\bf p}) d^3{\bf
p} \right)^N = 1.
\end{equation}

We consider the gas in the thermodynamic equilibrium with the
homogeneous density of particles $n({\bf r}) = n$. The density $n$
is expressed through the creation and annihilation operators in
coordinate representation, $a^{\dag}({\bf r})$ and $a({\bf r})$,
as follows:
\begin{equation}
n = \frac{N}{V} = {\rm tr} \left[ \varrho_{E} a^{\dag}({\bf r})
a({\bf r}) \right] = \ave{a^{\dag}({\bf r}) a({\bf r})}.
\end{equation}

In the momentum representation, we have
\begin{eqnarray} \label{a-dag-a-average}
\ave{a^{\dag}({\bf p}) a({\bf p}')} &=& {\rm tr} \left[
\varrho_{E} a^{\dag}({\bf p}) a({\bf p}') \right]
\nonumber\\
&=& (2\pi\hbar)^3 n f({\bf p}) \delta({\bf p} - {\bf p}'),
\end{eqnarray}
where $\delta$ is the Dirac delta function (in this case, in a
three-dimensional space of momenta).

The Hamiltonian of a single gas particle is $H_1 = \int \frac{{\bf
p}^2}{2m} \ket{{\bf p}} \bra{{\bf p}} d^3{\bf p}$. Its second
quantization gives the environment Hamiltonian
\begin{equation}
H_{E} = \int \frac{{\bf p}^2}{2m} a^{\dag}({\bf p}) a({\bf p})
d^3{\bf p}.
\end{equation}

Let $H_S = \sum_k \epsilon_k \ket{k}\bra{k}$ be the system
Hamiltonian and $H_{S1}$ be the interaction Hamiltonian for the
system and a single gas particle. The total interaction
Hamiltonian $H_{\rm int}$ is the second quantization of $H_{S1}$.
For instance, if $H_{S1} = Q_S \otimes U({\bf r})$, then $H_{\rm
int} = Q_S \otimes \int U({\bf r}) a^{\dag}({\bf r}) a({\bf r})
d^3{\bf r}$.

The system and the gas environment altogether evolve in accordance
with the von Neumann equation
\begin{equation}
\frac{d \varrho_{S+E}}{dt} = -\frac{i}{\hbar}[ H_S \otimes I_{E} +
I_S \otimes H_{E} + H_{\rm int},\varrho_{S+E}]
\end{equation}
with the initial condition $\varrho_{S+E}(0) =
\varrho_S(0) \otimes \varrho_{E}$. The reduced system evolution is
obtained by taking the partial trace over environment,
\begin{equation} \label{system-general}
\frac{d \varrho_{S}}{dt} = {\rm tr}_{E} \left( -\frac{i}{\hbar}[
H_S \otimes I_{E} + I_S \otimes H_{E} + H_{\rm
int},\varrho_{S+{E}}] \right).
\end{equation}

The fundamental result of the LDL approach~\cite{dumcke-1985} is
that the open dynamics~\eqref{system-general} in the limit $n
\rightarrow 0$, $t \rightarrow +\infty$, $nt = {\rm const.}$,
reduces to Eq.~\eqref{master-equation} with the GKSL
generator~\eqref{GKSL}, namely,
\begin{equation}
\frac{d\varrho_S}{dt} = - \frac{i}{\hbar} [H_S + H_{\rm
LS},\varrho_S] + {\cal D}[\varrho_S].
\end{equation}
Importantly, the Lamb shift $H_{\rm LS}$ and the dissipator ${\cal
D}$ depend only on the scattering $\hat{T}$-operator for the
interaction of the system with one particle of the gas,
\begin{eqnarray}
\hat{T} = H_{S1}  \lim_{t \rightarrow \infty} \!\!\! & \bigg\{ &
\!\!\! \exp\left[ - \frac{it}{\hbar} \biggl( H_S \otimes I_{1} +
I_S
\otimes H_1 + H_{S1} \biggr) \right] \nonumber\\
&& \!\!\! \times \exp\left[ \frac{it}{\hbar} \biggl( H_S \otimes
I_{1} + I_S \otimes H_1 \biggr) \right] \bigg\}.
\end{eqnarray}
Denoting $T(k,{\bf q}|l,{\bf p}) := \bra{k} \otimes
\bra{\bf q} \hat{T} \ket{l} \otimes \ket{\bf p}$ and
\begin{equation}
T_{\epsilon}({\bf q},{\bf p}) = \sum_{k,l:~\epsilon_k - \epsilon_l
= \epsilon} T(k,{\bf q}|l,{\bf p}) \ket{k} \bra{l},
\end{equation}
the final result is~\cite{dumcke-1985}
\begin{eqnarray}
H_{\rm LS} &=& (2\pi\hbar)^3 n \sum_{k,l:\
\epsilon_k = \epsilon_l} \int d^3{\bf p} \, f({\bf p}) \,
{\rm Re} T(k,{\bf p}|l,{\bf p}) \, \ket{k}\bra{l},\quad\,\,\,\\
{\cal D}[\varrho_S] &=& (2\pi)^4 \hbar^2 n
\sum_{\epsilon} \iint d^3{\bf p} \, d^3{\bf q} \, f({\bf p}) \,
\delta\left( \frac{{\bf q}^2}{2m} -
\frac{{\bf p}^2}{2m} + \epsilon \right) \nonumber\\
&\times& \left[ T_{\epsilon}({\bf q},{\bf
p}) \varrho_S T_{\epsilon}^{\dag}({\bf q},{\bf p}) - \frac{1}{2}
\left\{ \varrho_S, T_{\epsilon}^{\dag}({\bf q},{\bf p})
T_{\epsilon}({\bf q},{\bf p}) \right\} \right].
\end{eqnarray}
Here, we restored the physical dimension of the Lamb shift
(energy) and the dissipator (frequency) and have taken into
account the factor $(2\pi\hbar)^3$ from
Eq.~\eqref{a-dag-a-average}.

In what follows, we consider a modification of the LDL approach
for the case of gas particles having also internal degrees of
freedom, e.g., spin.

\subsection{Gas of particles with internal degrees of freedom}

Let $\{\ket{i}\}_i$ be an eigenbasis for the internal Hamiltonian
of gas particles, $H_{\lambda} = \sum_i \lambda_i \ket{i}\bra{i}$.
Merging the motional and internal degrees of freedom in the
notation $\ket{i,{\bf p}}$, we denote the corresponding creation
and annihilation operators by
\begin{equation}
a_i^{\dag}({\bf p}) := a^{\dag}(i, {\bf p}), \quad a_i({\bf p}) :=
a(i, {\bf p}).
\end{equation}

Suppose that the internal state of every gas particle is $\sum_{i}
\mu_i \ket{i} \bra{i}$. Then the environmental state is
$\widetilde{\varrho}_{E} = \widetilde{\varrho}_1^{\otimes N}$ with
\begin{equation} \label{rho-env-modified}
\widetilde{\varrho}_1 = \frac{(2\pi\hbar)^3}{V} \sum_{i} \mu_i
\int f({\bf p}) \ket{i,{\bf p}} \bra{i,{\bf p}} d^3{\bf p}.
\end{equation}

The single-particle Hamiltonian $\widetilde{H}_1 := H_{\lambda}
\otimes I_1 + I_{\lambda} \otimes H_1$ represents the sum of
internal and kinetic energies, respectively. The second quantized
version of $\widetilde{H}_1$ is
\begin{equation}
\widetilde{H}_{E} = \sum_{i} \int d^3{\bf p} \left( \lambda_i +
\frac{ {\bf p}^2}{2m} \right) a_i^{\dag}({\bf p}) a_i({\bf p})
\end{equation}
and commutes with $\widetilde{\varrho}_{E}$.

\begin{figure}
\centering
\includegraphics[width=8cm]{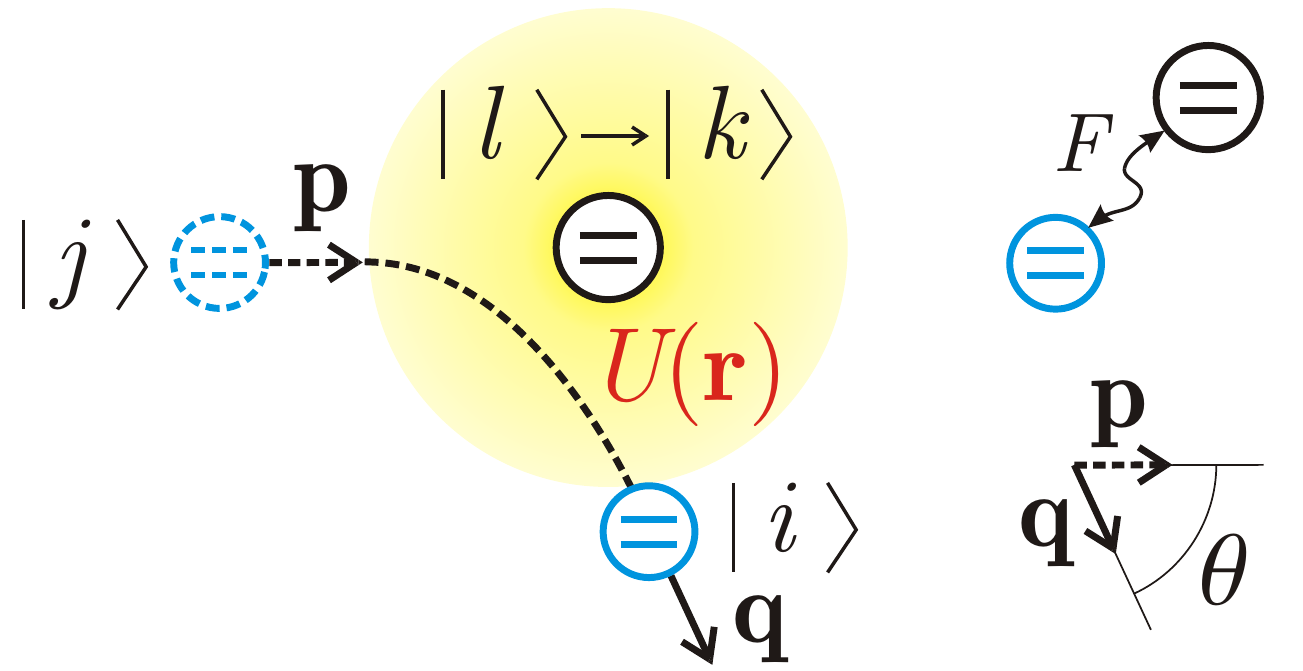}
\caption{A gas particle with the initial momentum ${\bf p}$ and
internal state $\ket{j}$ is scattered to the state with momentum
${\bf q}$ and internal state $\ket{i}$, whereas the system state
is changed from $\ket{l}$ to $\ket{k}$. Operator $F$ defines the
interaction between internal degrees of freedom of the gas
particle and the system, potential $U({\bf r})$ determines the
strength of the interaction.} \label{figure2}
\end{figure}

This model allows for including the interaction between the system and the
internal degrees of freedom of gas particles during collisions. We
consider the interaction Hamiltonian of the form
\begin{equation} \label{F-U}
\widetilde{H}_{S1} = F \otimes U({\bf r}) =  \sum_{k,l,i,j}
F_{ki,lj} \ket{k} \bra{l} \otimes \ket{i}\bra{j} \otimes U({\bf
r}),
\end{equation}
where the operator $F$ describes interaction between internal
degrees of freedom of the system and a gas particle, and $U({\bf
r})$ determines the strength of this interaction for a given
position ${\bf r}$ of the gas particle with respect to the system,
see Fig.~\ref{figure2}.

The scattering operator for this model is
\begin{eqnarray}
\widetilde{T} = \widetilde{H}_{S1}  \lim_{t \rightarrow \infty}
\!\!\! & \bigg\{ & \!\!\! \exp\left[ - \frac{it}{\hbar} \left( H_S
\otimes \widetilde{I}_{1} + I_S
\otimes \widetilde{H}_1 + \widetilde{H}_{S1} \right) \right] \nonumber\\
&& \!\!\! \times \exp\left[ \frac{it}{\hbar} \left( H_S \otimes
\widetilde{I}_{1} + I_S \otimes \widetilde{H}_1 \right) \right]
\bigg\},
\end{eqnarray}
where $\widetilde{I}_{1}$ is the identity operator for
the gas particle. Denoting $\widetilde{T}(k;i,{\bf q}|l;j,{\bf p})
:= \bra{k;i,{\bf q}} \widetilde{T} \ket{l;j,{\bf p}}$ and
\begin{equation}
\widetilde{T}_{\epsilon}(i,{\bf q};j,{\bf p}) =
\sum_{k,l:~\epsilon_k - \epsilon_l = \epsilon}
\widetilde{T}(k;i,{\bf q}|l;j,{\bf p}) \ket{k} \bra{l},
\end{equation}
the final result for the Lamb shift and dissipator in
the LDL master equation is
\begin{eqnarray}
\widetilde{H}_{\rm LS} &=& (2\pi\hbar)^3 n
\sum_i  \sum_{k,l:\, \epsilon_k = \epsilon_l}  \mu_i\nonumber\\
&& \times \int
d^3{\bf p} \, f({\bf p}) \, {\rm Re} \widetilde{T}(k;i,{\bf p}|l;i,{\bf p}) \, \ket{k}\bra{l}, \label{LS-internal} \\
\widetilde{\cal D}[\varrho_S] &=& (2\pi)^4
\hbar^2 n \sum_{\epsilon} \sum_{i,j} \mu_j \iint d^3{\bf p} \, d^3{\bf q} \, f({\bf
p}) \delta\bigg( \frac{{\bf q}^2}{2m} + \lambda_i \nonumber\\
&& - \frac{{\bf p}^2}{2m} - \lambda_j + \epsilon \bigg) \bigg[
\widetilde{T}_{\epsilon}(i,{\bf q};j,{\bf p}) \varrho_S
\widetilde{T}_{\epsilon}^{\dag}(i,{\bf q};j,{\bf p}) \nonumber\\
&&- \frac{1}{2} \left\{ \varrho_S,
\widetilde{T}_{\epsilon}^{\dag}(i,{\bf q};j,{\bf p})
\widetilde{T}_{\epsilon}(i,{\bf q};j,{\bf p}) \right\} \bigg].
\qquad \label{D-internal}
\end{eqnarray}

\subsection{Gas of spin particles in the Born approximation}

Consider a gas of particles with degenerate internal degrees of
freedom, e.g., spin particles in zero magnetic field. In this
case, $\lambda_i = 0$ for all $i$ and $H_{\lambda} = 0$. To
further simplify the expression~\eqref{D-internal}, let us also
assume that the separation of system energy levels is small as
compared to the characteristic kinetic energy of gas particles,
i.e. that $|\epsilon_k - \epsilon_l| \ll \left\langle \frac{{\bf
p}^2}{2m} \right\rangle$. For instance, this holds if the system
is a spin in zero magnetic field. In this case, the collisions are
\emph{elastic} meaning that the energy of incident particles
equals the energy of scattered particles. Then $\epsilon$ takes
the only zero value, and we simplify the summations: $\sum_{k,l:\,
\epsilon_k = \epsilon_l} = \sum_{k,l}$ and
$\widetilde{T}_{0}(i,{\bf q};j,{\bf p}) = \sum_{k,l}
\widetilde{T}(k;i,{\bf q}|l;j,{\bf p}) \ket{k} \bra{l}$.
Additionally, we have
\begin{equation}
\delta\left( \frac{{\bf q}^2}{2m} - \frac{{\bf p}^2}{2m} \right) =
\frac{m}{p} \, \delta(q-p),
\end{equation}
where we use the notations $q=|{\bf q}|$ and $p = |{\bf p}|$.

To calculate the elements of the $T$-matrix analytically, we
consider the first-order Born approximation $\widetilde{T} \approx
\widetilde{H}_{S1}$ leading to
\begin{eqnarray} \label{T-Born}
\widetilde{T}(k;i,{\bf q}|l;j,{\bf p}) & \approx &  F_{ki,lj} \bra{\bf q} U({\bf r}) \ket{\bf p} \nonumber\\
& = & \frac{F_{ki,lj}}{(2\pi\hbar)^3} \int e^{i({\bf p} - {\bf
q}){\bf r} / \hbar} U({\bf r}) d^3{\bf r}. \quad
\end{eqnarray}

Let $U_0$ be the characteristic strength of $U({\bf r})$ and $d$
be the characteristic distance such that $U({\bf r})$ is
negligible if $|{\bf r}| > d$. Then the first-order Born
approximation is valid for fast particles with $pd \gg \hbar$ if
$U_0 \ll \frac{\hbar p}{m d}$~\cite{LL}. Since the average
momentum is $\ave{p} = \int |{\bf p}| f({\bf p}) d^3{\bf p} =
\sqrt{8mkT/\pi}$, the first-order Born approximation is valid for
fast particles if
\begin{equation} \label{Born}
U_0 \ll \sqrt{\frac{\hbar^2 kT}{m d^2}}.
\end{equation}

In the first-order Born approximation, substituting
Eq.~\eqref{T-Born} into the Lamb shift~\eqref{LS-internal} and the
dissipator~\eqref{D-internal} yields
\begin{eqnarray}
\widetilde{H}_{\rm LS}^{\rm LDL} &=& n \int
U({\bf r}) d^3{\bf r} \sum_{i}
\mu_i  A_{ii},  \label{Lamb-LDL} \\
\widetilde{\cal D}^{\rm LDL}[\varrho_S] &=&
\Gamma \sum_{i,j} \mu_j \left( A_{i j} \varrho_S A_{i j}^{\dag} -
\frac{1}{2} \left\{ \varrho_S, A_{i j}^{\dag} A_{i j} \right\}
\right), \qquad \label{D-LDL}
\end{eqnarray}

\noindent where we introduced the notations
\begin{eqnarray}
\label{A-definition} A_{i j} &=& \sum_{k,l}
F_{ki,lj} \ket{k} \bra{l} = I_S \otimes \bra{i} \, F \, I_S
\otimes \ket{j}, \\
\label{D-0} \Gamma &=& (2\pi)^4 \hbar^2 n m
\iint d^3{\bf p} \, d^3{\bf q}  \frac{\, f({\bf p}) \, \big\vert
\! \bra{\bf q} U({\bf r})
\ket{\bf p} \! \big\vert^2 \, \delta(q-p)}{p} \nonumber\\
\end{eqnarray}

\noindent and have taken into account $\int d^3{\bf p} f({\bf p})
= 1$.

Provided the potential $U({\bf r})$ is spherically symmetrical,
i.e., $U({\bf r}) = V(r)$, $r = |{\bf r}|$, the
expression~\eqref{D-0} can be further simplified. In this case,
the Fourier transform $\bra{\bf q} U({\bf r}) \ket{\bf p}$ depends
only on the absolute value $|{\bf q} - {\bf p}|$, which in turn
depends on the scattering angle $\theta$ between ${\bf p}$ and
${\bf q}$. Due to the presence of delta function $\delta(p-q)$ in
$\Gamma$, one can set $q=p$ that gives $|{\bf q} - {\bf p}| = 2 p
\sin \frac{\theta}{2}$ and
\begin{eqnarray}
&& \bra{\bf q} U({\bf r}) \ket{\bf p} \Big\vert_{q=p} \nonumber\\
&& = \frac{1}{(2\pi\hbar)^2 p \sin\frac{\theta}{2}}
\int\limits_0^{\infty} V(r) \sin\left( \frac{2pr}{\hbar}
\sin\frac{\theta}{2} \right) r dr.
\end{eqnarray}

Remembering that the distribution $f({\bf p})$ depends on the
absolute value of momentum $p = |{\bf p}|$, we further use the
notation $f(p)$ instead of $f({\bf p})$ to refer to
Eq.~\eqref{p-distribution}. This allows us to first integrate over
$d^3{\bf q} = q^2 dq  \sin\theta d\theta d\varphi$ and later use
the simplified expression $d^3{\bf p} = 4\pi p^2 dp$. Introducing
a new variable, $\xi = \sin\frac{\theta}{2}$, we have $\sin\theta
d\theta = 4\xi d\xi$ and finally
\begin{equation} \label{D-0-through-V}
\Gamma = \frac{32 \pi^2 n m}{\hbar^2} \int\limits_0^{\infty} f(p)
\, p \, dp \int\limits_0^1 \frac{d\xi}{\xi} \left(
\int\limits_0^{\infty} V(r) \sin \frac{2pr\xi}{\hbar} \, r dr
\right)^2.
\end{equation}

In what follows, we consider the particular cases of analytically
tractable potentials $V(r)$ to get the final explicit expression
for the dissipator $\widetilde{\cal D}^{\rm LDL}$.

\subsubsection{Gaussian potential}

Consider Gaussian potential $U({\bf r}) = V(r) = U_0 \exp\left(-
\dfrac{r^2}{2d^2} \right)$. Direct computation yields
\begin{equation} \label{scattering-amp-Gaussian}
\int\limits_0^{\infty} V(r) \sin \frac{2pr\xi}{\hbar} \, r dr =
\frac{\sqrt{2\pi} p d^3 U_0 \xi}{\hbar} \exp \left( - \frac{2 p^2
d^2 \xi^2}{\hbar^2}\right).
\end{equation}

\noindent Substituting Eq.~\eqref{scattering-amp-Gaussian} into
Eq.~\eqref{D-0-through-V}, we get
\begin{equation} \label{D-0-Gaussian}
\Gamma = \frac{(2\pi)^{3/2} n m d^4 U_0^2}{\hbar^2 \sqrt{mkT}
\left( 1+ \dfrac{\hbar^2}{8m d^2 kT}\right)}.
\end{equation}

Since the average momentum $\ave{p} = \sqrt{8mkT/\pi}$ satisfies
the condition $\ave{p}d \gg \hbar$ for fast particles, we neglect
the term $\frac{\hbar^2}{8m d^2 kT}$ in Eq.~\eqref{D-0-Gaussian}
and obtain
\begin{equation} \label{D-0-Gaussian-fast}
\Gamma \Big\vert_{\rm fast} = \frac{(2\pi)^{3/2} n m d^4
U_0^2}{\hbar^2 \sqrt{mkT}}.
\end{equation}

\noindent The derived expression is valid if the
condition~\eqref{Born} is additionally satisfied.

\subsubsection{Spherical square-well potential}

Consider the spherical square-well potential $U({\bf r}) = V(r) =
\left\{
\begin{array}{ll}
  U_0, & r \leq d, \\
  0, & r > d. \\
\end{array} \right.$ Then
\begin{equation} \label{scattering-amp-rectangular}
\int\limits_0^{\infty} V(r) \sin \frac{2pr\xi}{\hbar} \, r dr =
\frac{\hbar d U_0}{2p\xi} \left( \frac{\hbar}{2 p d \xi}
\sin\frac{2pd\xi}{\hbar} - \cos\frac{2pd\xi}{\hbar} \right).
\end{equation}

\noindent Substituting Eq.~\eqref{scattering-amp-rectangular} into
Eq.~\eqref{D-0-through-V}, we get a rather complicated expression,
which is simplified for fast particles with $\ave{p}d \gg \hbar$
as follows:
\begin{equation} \label{D-0-rectangular-fast}
\Gamma \Big\vert_{\rm fast} = \frac{2\sqrt{2\pi} n m d^4
U_0^2}{\hbar^2 \sqrt{mkT}}.
\end{equation}

\noindent Note that the obtained result is derived within the
first-order Born approximation that is valid if the
condition~\eqref{Born} is satisfied.

\section{Semiclassical collision model}
\label{section-cm}

\subsection{Collision model with a finite interaction time}
\label{section-cm-finite-int-time}

\begin{figure}
\centering
\includegraphics[width=8cm]{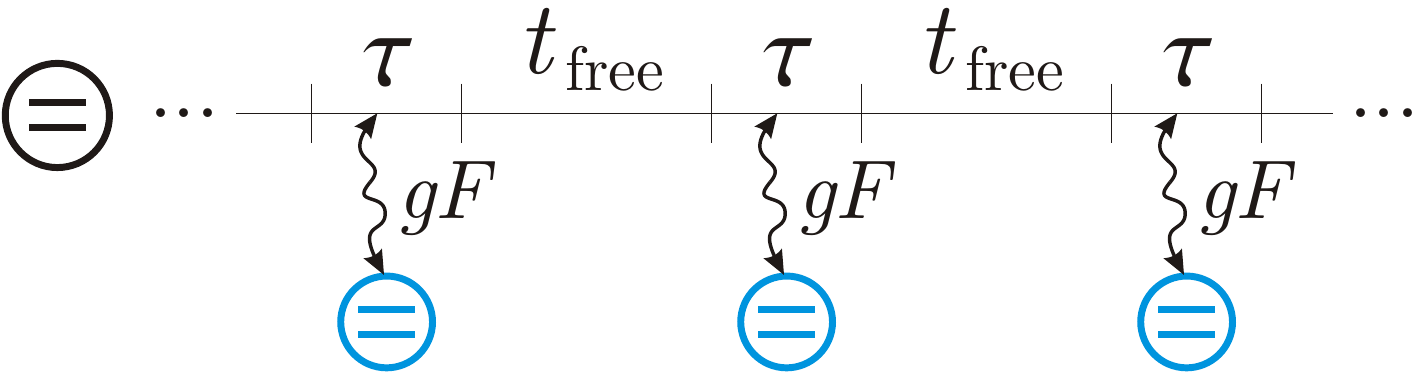}
\caption{Collision model with impact time $\tau$ and free
propagation time $t_{\rm free}$. The system-particle Hamiltonian
during the collision is $gF$.} \label{figure3}
\end{figure}

In conventional collision models~\cite{rau-1963,scarani-2002}, the
quantum system sequentially interacts with environmental particles,
whose only degrees of freedom are internal. The system interacts
with each environmental particle only once, and the initial state of
all environment particles is $\left(\sum_i \mu_i
\ket{i}\bra{i}\right)^{\otimes N}$. Each collision lasts for a
finite time $\tau$. In between the collisions, the system evolves
unitarily with its Hamiltonian $H_S$. Denote by $t_{\rm free}$ the
intercollision time. Then the frequency of collisions equals
$(t_{\rm free} + \tau)^{-1}$, see Fig.~\ref{figure3}.

Let $gF$ be the system-particle Hamiltonian during the collision,
where $g$ is the characteristic strength. This implies that one
can neglect the effect of the system Hamiltonian during the
collision, which is justified if $\tau \| [H_S \otimes I,F] \| \ll
\hbar \| F \|$. In particular, it takes place if $|\epsilon_k -
\epsilon_l| \tau \ll \hbar$. Assuming $g\tau \ll \hbar$, we obtain
the following master equation for the system:
\begin{eqnarray} \label{master-eq-collision}
&& \!\!\!\!\!\!\!\!\!\! \frac{d\varrho_S}{dt} = -
\frac{i}{\hbar(t_{\rm free} + \tau)} \left[ t_{\rm free} H_S + g
\tau \sum_{i}
\mu_i A_{ii}, \varrho_S \right] \nonumber\\
&& \!\!\!\!\!\!\!\!\!\! + \frac{g^2 \tau^2}{\hbar^2 (t_{\rm free}
+ \tau)} \sum_{i,j} \mu_j \left( A_{i j} \varrho_S A_{i j}^{\dag}
- \frac{1}{2} \left\{ \varrho_S, A_{i j}^{\dag} A_{i j} \right\}
\right), \qquad
\end{eqnarray}

\noindent where the operators $A_{ij}$ are expressed through $F$
exactly as in Eq.~\eqref{A-definition}.

If $\tau \gg t_{\rm free}$, then the obtained master equation is
valid in the limit $g \tau \rightarrow 0$, $g^2 \tau \rightarrow
{\rm const.}$~\cite{giovannetti-2012,lorenzo-2017,luchnikov-2017}.
If $\tau \ll t_{\rm free}$, then Eq.~\eqref{master-eq-collision}
reduces to
\begin{eqnarray} \label{master-eq-collision-simplified}
&& \frac{d\varrho_S}{dt} = - \frac{i}{\hbar} \left[ H_S + \frac{g
\tau}{t_{\rm free}} \sum_{i}
\mu_i A_{ii}, \varrho_S \right] \nonumber\\
&& + \frac{g^2 \tau^2}{\hbar^2 t_{\rm free}} \sum_{i,j} \mu_j
\left( A_{i j} \varrho_S A_{i j}^{\dag} - \frac{1}{2} \left\{
\varrho_S, A_{i j}^{\dag} A_{i j} \right\} \right), \qquad
\end{eqnarray}

\noindent and is valid if $g\tau \ll \hbar$.

Finally, consider an ensemble of particles with various values of
the parameter $g\tau$ that appear with various frequencies $t_{\rm
free}^{-1}$. Collisions with such an ensemble result in the Lamb
shift and the dissipator as follows:
\begin{eqnarray}
&& \!\!\!\!\!\!\!\!\!\! H_{\rm LS} = \left\langle \frac{g
\tau}{t_{\rm free}}
\right\rangle \sum_{i} \mu_i A_{ii}, \\
&& \!\!\!\!\!\!\!\!\!\! {\cal D}[\varrho_S] = \left\langle
\frac{g^2 \tau^2}{\hbar^2 t_{\rm free}} \right\rangle \sum_{i,j}
\mu_j \left( A_{i j} \varrho_S A_{i j}^{\dag} - \frac{1}{2}
\left\{ \varrho_S, A_{i
j}^{\dag} A_{i j} \right\} \right). \nonumber\\
\end{eqnarray}

\subsection{Collision model for gas particles}\label{section-finite-time}

In the semiclassical collision model, gas particles move along the
classical trajectories, whereas their internal degrees of freedom
are quantum. We consider a low density gas ($n d^3 \ll 1$), so
that the collisions are rather rare and we can neglect the events
when two or more gas particles are simultaneously in the volume
$\sim d^3$ nearby the system. It means that the effective
interaction time $\tau$ is much less than the intercollision time
$t_{\rm free}$.

Consider an itinerant gas particle with the given trajectory ${\bf
r}(t)$ that moves in the potential $U({\bf r})$ with
characteristic length $d$. Define the effective collision time
$\tau$ through
\begin{equation} \label{effective-time}
U_0 \tau = \int_{-\infty}^{+\infty} U \big({\bf r}(t) \big) dt,
\end{equation}

\noindent where $U_0$ is the characteristic strength of the
potential $U({\bf r})$. Then a single collision with the
interaction Hamiltonian~\eqref{F-U} results in the unitary
operator $W = \exp(-\frac{i}{\hbar} U_0 F \tau)$ that acts on the
internal degrees of freedom of the system and the itinerant gas
particle. Therefore, $U_0 \tau$ plays the same role as $g\tau$ in
Sec.~\ref{section-cm-finite-int-time}. Note that despite the fact
that a particle enters the interaction region $|{\bf r}| < d$ for
a finite period $(t_{\rm in},t_{\rm out})$, we can still use
definition~\eqref{effective-time} because the potential $U({\bf
r})$ is negligible when a gas particle is outside the interaction
region.

\begin{figure}
\centering
\includegraphics[width=8cm]{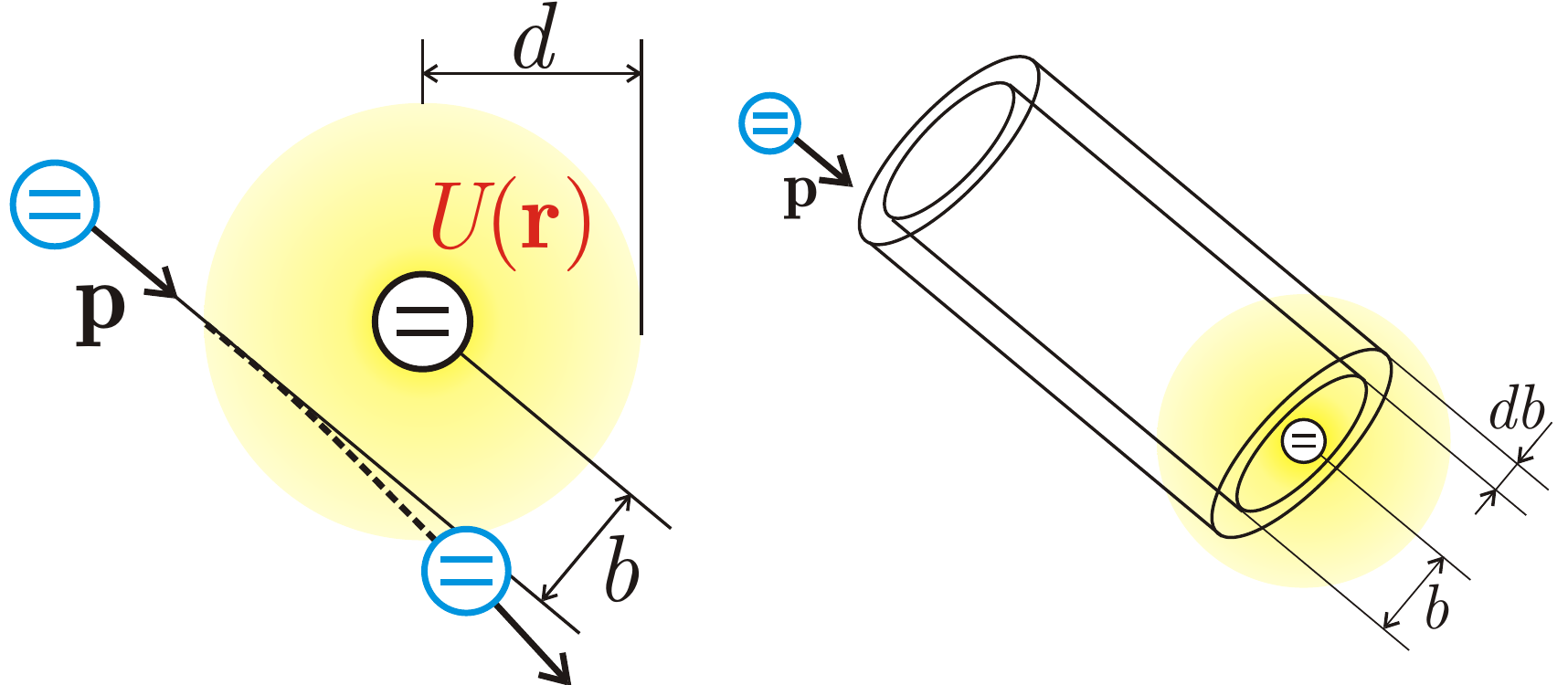}
\caption{The impact parameter $b$. The classical trajectories are
approximated by straight lines for fast gas particles (left). The
volume of particles with momentum ${\bf p}$ and impact parameter
$b - (b+db)$ that reach the interaction region within time $t$, is
$dV = 2\pi b \, db \times p t/ m$ (right).} \label{figure4}
\end{figure}

If the interaction strength between the system and a particle
($\sim U_0$) is small as compared to the kinetic energy of a gas
particle ($\sim kT$), then we can neglect the curvature of
trajectories and approximate them by straight lines, see
Fig.~\ref{figure4}. As before, we additionally assume that $U({\bf
r}) = V(r)$, i.e., the potential is spherically symmetrical.
Within such an approximation, $U_0 \tau$ depends on the absolute
value of particle momentum, $p$, and the impact parameter $b$ (see
Fig.~\ref{figure4}):
\begin{equation} \label{effective-time-simplified}
U_0 \tau = \int_{-\infty}^{+\infty} V \left( \sqrt{b^2 +
\tfrac{p^2 t^2}{m^2}} \right) dt.
\end{equation}

Consider particles with momenta ${\bf p} - ({\bf p} + d{\bf p})$.
The number of particles that would pass through the interaction
region with impact parameters $b - (b + db)$ within time period
$t$ equals $n dV f({\bf p}) d^3{\bf p}$, where $dV = 2\pi b \, db
\times p t/ m$ is the corresponding volume, see
Fig.~\ref{figure4}. Therefore, the collision rate for such
particles reads
\begin{equation}
\frac{1}{t_{\rm free}} = \frac{n \times 2\pi b \, db \times p
f({\bf p}) d^3{\bf p}}{m}  = \frac{ 8 \pi^2 n b p^3 f(p) \, db \,
dp }{ m }.
\end{equation}

Using the results of Sec.~\ref{section-cm-finite-int-time}, we
readily find the Lamb shift and the dissipator in the
semiclassical collision model:
\begin{eqnarray}
&& H_{\rm LS}^{\rm CM} = \left\langle \frac{U_0 \tau}{t_{\rm free}} \right\rangle \sum_{i} \mu_i A_{ii},  \label{Lamb-CM} \\
&& {\cal D}^{\rm CM} = \left\langle \frac{U_0^2 \tau^2}{\hbar^2
t_{\rm free}} \right\rangle \sum_{i,j} \mu_j \left( A_{i j}
\varrho_S A_{i j}^{\dag} - \frac{1}{2} \left\{ \varrho_S, A_{i
j}^{\dag} A_{i j} \right\} \right). \label{D-CM} \nonumber\\
\end{eqnarray}

\noindent Here
\begin{eqnarray}
&& \label{C1} \left\langle \frac{U_0 \tau}{t_{\rm free}}
\right\rangle = \int_0^{\infty} db \int_0^{\infty} dp \frac{ 8
\pi^2 n b p^3
f(p)}{ m } \nonumber\\
&& \qquad \quad \times \int_{-\infty}^{+\infty} V \left( \sqrt{b^2
+ \tfrac{p^2
t^2}{m^2}} \right) dt ,\\
&& \label{C2} \left\langle \frac{U_0^2 \tau^2}{\hbar^2 t_{\rm
free}} \right\rangle = \int_0^{\infty} db \int_0^{\infty} dp
\frac{ 8 \pi^2 n b p^3 f(p)}{ \hbar^2 m }
\nonumber\\
&& \qquad \quad \times \left[\int_{-\infty}^{+\infty} V \left(
\sqrt{b^2 + \tfrac{p^2 t^2}{m^2}} \right) dt \right]^2.
\end{eqnarray}

Since $\tau \sim \frac{md}{\ave{p}} \sim \sqrt{\frac{md^2}{kT}}$
and $t_{\rm free} \sim \frac{m}{n d^2 \ave{p}}$, the derived
formulas are valid if $nd^3 \ll 1$ (approximation of rare
collisions, $\tau \ll t_{\rm free}$), $|\epsilon_k - \epsilon_l|
\sqrt{\frac{md^2}{kT}} \ll \hbar$ and $U_0 \sqrt{\frac{md^2}{kT}}
\ll \hbar$ (stroboscopic approximation), $kT \gg U_0$
(approximation of straight trajectories).

In what follows, we consider particular cases of analytically
tractable potentials $V(r)$ to get the explicit expressions for
Eqs.~\eqref{C1} and \eqref{C2}.

\subsubsection{Gaussian potential}

If $U({\bf r}) = V(r) = U_0 \exp\left(- \dfrac{r^2}{2d^2}
\right)$, then
\begin{equation} \label{int-V-Gaussian}
\int_{-\infty}^{+\infty} V \left( \sqrt{b^2 + \tfrac{p^2
t^2}{m^2}} \right) dt = \frac{\sqrt{2\pi} m d U_0}{p} \exp\left(-
\frac{b^2}{2 d^2}\right).
\end{equation}

\noindent Substituting Eq.~\eqref{int-V-Gaussian} into
Eqs.~\eqref{C1} and \eqref{C2}, we get
\begin{eqnarray}
&& \label{C1-Gaussian} \left\langle \frac{U_0 \tau}{t_{\rm free}}
\right\rangle = (2\pi)^{3/2} nd^3 U_0,\\
&& \label{C2-Gaussian} \left\langle \frac{U_0^2 \tau^2}{\hbar^2
t_{\rm free}} \right\rangle = \frac{(2\pi)^{3/2} n m d^4
U_0^2}{\hbar^2 \sqrt{mkT}}.
\end{eqnarray}

\subsubsection{Spherical square-well potential}

If $U({\bf r}) = V(r) = \left\{ \begin{array}{ll}
  U_0, & r \leq d, \\
  0, & r > d, \\
\end{array} \right.$ then
\begin{equation} \label{int-V-rectangular}
\int_{-\infty}^{+\infty} V \left( \sqrt{b^2 + \tfrac{p^2
t^2}{m^2}} \right) dt = \left\{ \begin{array}{ll}
  \frac{2mU_0}{p} \sqrt{d^2 - b^2}, & b \leq d, \\
  0, & b > d. \\
\end{array} \right.
\end{equation}

\noindent Substituting Eq.~\eqref{int-V-rectangular} into
Eqs.~\eqref{C1} and \eqref{C2}, we get
\begin{eqnarray}
&& \label{C1-rectangular} \left\langle \frac{U_0 \tau}{t_{\rm
free}}
\right\rangle = \frac{4\pi}{3} nd^3 U_0,\\
&& \label{C2-rectangular} \left\langle \frac{U_0^2 \tau^2}{\hbar^2
t_{\rm free}} \right\rangle = \frac{2\sqrt{2\pi} n m d^4
U_0^2}{\hbar^2 \sqrt{mkT}}.
\end{eqnarray}

\section{Comparison of the two approaches}
\label{section-comparison}

\subsection{Comparison in the high temperature limit}

In Secs.~\ref{section-ldl} and \ref{section-cm}, the two different
approaches are presented for the derivation of the GKSL master
equation for a spin system interacting with a diluted gas of spin
particles. In the low density limit of the fully quantum approach,
the generator of the master equation is defined by
formulas~\eqref{Lamb-LDL} and \eqref{D-LDL}. In the semiclassical
collision model, the generator of the master equation is defined
by formulas~\eqref{Lamb-CM} and \eqref{D-CM}.

The first observation is that both generators are expressed
through the same operators $A_{ij}$ and have identical operator
structure.

Second, the Lamb shifts~\eqref{Lamb-LDL} and \eqref{Lamb-CM}
exactly coincide because by the change of variables $z = pt/m$ in
Eq.~\eqref{C1} we extract $\int_0^{\infty} 4 \pi^2 p^2 f(p) dp =1$
and get the following integral in cylindrical coordinates:
\begin{eqnarray}
\left\langle \frac{U_0 \tau}{t_{\rm free}} \right\rangle &=& n
\int_0^{\infty} 2 \pi b db \int_{-\infty}^{+\infty} V \left(
\sqrt{b^2
+ z^2} \right) dz \nonumber\\
& = & n \int U({\bf r})d^3{\bf r}.
\end{eqnarray}

Third, the dissipators~\eqref{D-LDL} and \eqref{D-CM} generally do
not exactly coincide because $\Gamma \neq \left\langle \frac{U_0^2
\tau^2}{\hbar^2 t_{\rm free}} \right\rangle$ for finite
temperatures, cf. Eqs.~\eqref{D-0-Gaussian} and
\eqref{C2-Gaussian}. However, for the considered examples of
Gaussian and spherical square-well potentials surprisingly $\Gamma
\vert_{\rm fast} = \left\langle \frac{U_0^2 \tau^2}{\hbar^2 t_{\rm
free}} \right\rangle$. In fact, if the average kinetic energy $kT
\gg \frac{\hbar^2}{md^2}$, then the gas particles are fast and the
dominant scattering angles satisfy $\theta \lesssim
\frac{\hbar}{pd}$, Ref.~\cite{LL}. In this case, $\xi =
\sin\frac{\theta}{2} \lesssim \frac{\hbar}{2pd}$ and
\begin{eqnarray}
&& \int_0^1 \frac{d\xi}{\xi} \left( \int_0^{\infty}
V(r) \sin \frac{2pr\xi}{\hbar} \, r dr \right)^2 \nonumber\\
&& \approx \int_0^{\frac{\hbar}{2pd}} \frac{d\xi}{\xi} \left(
\int_0^d V(r) \sin \frac{2pr\xi}{\hbar} \, r dr \right)^2 \nonumber\\
&& \approx \int_0^{\frac{\hbar}{2pd}} \frac{d\xi}{\xi} \left(
\int_0^d V(r) \frac{2pr\xi}{\hbar} \, r dr \right)^2 \nonumber\\
&& = \frac{1}{2d^2} \left( \int_0^d V(r) r^2 dr \right)^2 \sim
U_0^2 d^4.
\end{eqnarray}

\noindent The obtained estimation is of the same order as the
collision model expression,
\begin{equation}
\int_0^{\infty} b \, db \left[ \int V(\sqrt{b^2 + z^2}) dz
\right]^2 \sim U_0^2 d^4.
\end{equation}

\noindent Therefore, $\Gamma \sim \left\langle \frac{U_0^2
\tau^2}{\hbar^2 t_{\rm free}} \right\rangle$ if $kT \gg
\frac{\hbar^2}{md^2}$.

Fourth, in the limit of infinite temperature the dissipators in
the low density approach and the collision model exactly coincide
for spherical potentials $U({\bf r})=V(r)$, i.e.,
\begin{equation} \label{T-limit}
\lim_{T \rightarrow \infty} \frac{\Gamma}{\left\langle \frac{U_0^2
\tau^2}{\hbar^2 t_{\rm free}} \right\rangle} = 1.
\end{equation}

\noindent To prove Eq.~\eqref{T-limit} we rewrite the integral
$\int_{-\infty}^{+\infty} V \left( \sqrt{b^2 + z^2} \right) dz = 2
\int_{b}^{\infty} V(r) \frac{r \, dr}{\sqrt{r^2 - b^2}}$ which
yields
\begin{eqnarray}
&& \int_0^{\infty} b \, db \left[ \int_{-\infty}^{+\infty} V
\left( \sqrt{b^2 + z^2} \right) dz \right]^2 \nonumber\\
&& = 4 \int\limits_0^{\infty} b \, db \int\limits_0^{\infty} dr
\int\limits_0^{\infty} dr' V(r) V(r') r r' f(b,r) f(b,r'), \qquad
\end{eqnarray}

\noindent where $f(b,r) = \left\{ \begin{array}{ll}
  0 & \text{if~} r \leq b, \\
  \frac{1}{\sqrt{r^2-b^2}} & \text{if~} r > b. \\
\end{array} \right.$ Since
\begin{eqnarray}
&& \int_0^{\infty} b \, db \, f(b,r) f(b,r') = \int_0^{\min(r,r')}
\frac{b \, db}{\sqrt{(r^2-b^2)(r'^2-b^2)}} \nonumber\\
&& = \frac{1}{2} \ln \frac{r+r'}{|r-r'|},
\end{eqnarray}

\noindent we get the following expression in the collision model:
\begin{eqnarray} \label{C2-kernel}
&& \left\langle \frac{U_0^2 \tau^2}{\hbar^2 t_{\rm free}}
\right\rangle = \frac{16 \pi^2 n m}{\hbar^2}
\int_0^{\infty} f(p) \, p \, dp \nonumber\\
&& \qquad \times \int_0^{\infty} dr \int_0^{\infty} dr' V(r) V(r')
r r' \ln \frac{r+r'}{|r-r'|}.
\end{eqnarray}

\noindent On the other hand, in the low density approach,
Eq.~\eqref{D-0-through-V} can be rewritten in the form
\begin{equation} \label{Gamma-kernel}
\Gamma = \frac{32 \pi^2 n m}{\hbar^2} \int\limits_0^{\infty} f(p)
\, p \, dp \int\limits_0^{\infty} dr \int\limits_0^{\infty} dr' \,
V(r) V(r') r r' K(r,r'),
\end{equation}

\noindent where the kernel
\begin{equation}
K(r,r') = \int\limits_0^1 \frac{d\xi}{\xi} \sin
\frac{2pr\xi}{\hbar} \sin \frac{2pr'\xi}{\hbar} \xrightarrow[]{p
\rightarrow \infty} \frac{1}{2} \ln \frac{r+r'}{|r-r'|}.
\end{equation}

\noindent As the limit $p \rightarrow \infty$ is equivalent to the
high temperature limit $T \rightarrow \infty$, we see that
Eqs.~\eqref{C2-kernel} and \eqref{Gamma-kernel} coincide when $T
\rightarrow \infty$, which leads to Eq.~\eqref{T-limit}.

Fifth, the applicability of the first-order Born approximation for
fast particles in the low-density-limit approach,
Eq.~\eqref{Born}, is equivalent to the condition of stroboscopic
approximation in the collision model, $g\tau \ll \hbar
\Leftrightarrow U_0 \sqrt{\frac{md^2}{kT}} \ll \hbar$.

Sixth, if both conditions $kT \gg \frac{\hbar^2}{md^2}$ (fast
particles) and $U_0 \sqrt{\frac{md^2}{kT}} \ll \hbar$ (Born
approximation and stroboscopic approximation) are satisfied, then
automatically $kT \gg U_0$, i.e., the approximation of straight
trajectories is justified in the collision model.

Finally, we conclude that both the low density limit approach and
the collision model provide very similar predictions for the
reduced dynamics of the spin system ($\epsilon_k = \epsilon_l$,
$\lambda_i = 0$) if $nd^3\ll 1$, $kT \gg \frac{\hbar^2}{md^2}$,
and $U_0 \ll \sqrt{\frac{\hbar^2 kT}{m d^2}}$.

\subsection{Estimation of difference for finite temperature}

Let us analyze the difference between the two approaches when
lowering the gas temperature. We consider a spherical square-well
potential, for which the quantitative estimation of the
discrepancy becomes tractable.

In the LDL approach, lowering the velocity of gas particles can be
taken into account by considering the second-order perturbation of
the scattering operator, $\widetilde{T} = F \otimes U({\bf r}) + F
\otimes U({\bf r}) G_0^{(+)}(E) F \otimes U({\bf r})$, where
$G_0^{(+)}(E)$ is the retarded Green operator for Hamiltonian $H_S
\otimes \widetilde{I}_{1} + I_S \otimes \widetilde{H}_1$,
$E=\frac{p^2}{2m}$. Provided $kT \gg \frac{\hbar^2}{md^2}$, we
find the matrix element $\widetilde{T}(k;i,{\bf p}|l;i,{\bf p})$
and calculate the corrected Lamb shift:
\begin{equation} \label{LS-corrected}
\widetilde{H}_{\rm LS}^{\rm LDL} = \frac{4\pi}{3} n d^3 U_0
\sum_{i,k,l} \mu_i \left( F_{ki,li} - \frac{2 U_0}{kT}
(F^2)_{ki,li} \right) \ket{k}\bra{l}.
\end{equation}

\noindent Finite values of $\frac{k T m d^2}{\hbar^2}$ lead to the
exponentially small relative error in the Lamb shift of the order
of $\frac{\hbar}{d \sqrt{mkT}} \exp(- \frac{k T m d^2}{\hbar^2})$
as a result of approximate integration
\begin{eqnarray}
&& \int_{|{\bf r}'| \leq d} d^3{\bf r}' \int d^3{\bf p} \, f({\bf
p}) \, \frac{e^{i |{\bf r} - {\bf r}'| p / \hbar}}{|{\bf r} - {\bf
r}'|}
\nonumber\\
&& \approx \int_{{\bf r}' \in \mathbb{R}^3} d^3{\bf r}' \int
d^3{\bf p} \, f({\bf p}) \, \frac{e^{i |{\bf r} - {\bf r}'| p /
\hbar}}{|{\bf r} - {\bf r}'|}.
\end{eqnarray}

\noindent We see that the Lamb shifts $\widetilde{H}_{\rm LS}^{\rm
LDL}$ and $H_{\rm LS}^{\rm CM} = \left\langle \frac{U_0
\tau}{t_{\rm free}} \right\rangle \sum_{i,k,l} \mu_i F_{ki,li}$
have different operator structure in general. If $\left\langle
\frac{U_0 \tau}{t_{\rm free}} \right\rangle$ is given by
Eq.~\eqref{C1-rectangular}, then the relative error
\begin{eqnarray} \label{H-H}
&& \frac{\| \widetilde{H}_{\rm LS}^{\rm LDL} - H_{\rm LS}^{\rm
CM}\|}{ n d^3 |U_0| \, \|F\|} \nonumber\\
&& \sim \max \left[ \frac{|U_0| \, \|F\|}{kT} , \frac{\hbar}{d
\sqrt{mkT}} \exp\left(- \frac{k T m d^2}{\hbar^2}\right) \right].
\end{eqnarray}

As far as the dissipator in the LDL approach is concerned, the
small parameter $\frac{\hbar^2}{k T m d^2}$ contributes linearly
already in the first-order Born approximation [cf.
Eq.~\eqref{D-0-Gaussian} for the Gaussian potential]. In fact, for
a spherical square-well potential we have
\begin{eqnarray}
&& \int_0^1 \frac{d\xi}{\xi} \left( \int_0^{\infty} V(r) \sin
\frac{2pr\xi}{\hbar} \, r dr \right)^2 = \frac{\hbar^4 U_0^2}{128
p^3} \nonumber\\
&& \times \left( \frac{32 p^4 d^4}{\hbar^4} - \frac{8 p^2
d^2}{\hbar^2} - 1 + \cos \frac{4pd}{\hbar} + \frac{4pd}{\hbar}
\sin \frac{4pd}{\hbar} \right) \nonumber\\
&& \approx \frac{U_0^2 d^4}{4} \left[ 1 - \exp\left( - \frac{8 p^2
d^2}{9 \hbar^2} \right) \right], \label{int-approx}
\end{eqnarray}

\noindent where the latter approximation provides an interpolation
between asymptotics $\frac{2 p^2 U_0^2 d^6}{9 \hbar^2}$ for
$\frac{pd}{\hbar} \ll 1$ and $\frac{U_0^2 d^4}{4}$ for
$\frac{pd}{\hbar} \gg 1$ and has the maximum relative error 4.21\%
for $\frac{pd}{\hbar} = 5.03$. Substituting~\eqref{int-approx}
into \eqref{D-0-through-V}, we get
\begin{equation}
\Gamma = \frac{2\sqrt{2\pi} n m d^4 U_0^2}{\hbar^2 \sqrt{mkT}}
\left( 1 - \frac{9 \hbar^2}{16 k T m d^2} \right).
\end{equation}

Similarly to the case of the Lamb shift, we expect that the second
order perturbation with respect to the small parameter
$\frac{|U_0|}{kT}$ in the LDL approach would result in the jump
operators that are different from the jump operators in the
collision model. Therefore, the relative discrepancy in
dissipators is estimated as
\begin{equation} \label{D-D}
\frac{\| \widetilde{\cal D}^{\rm LDL}[\varrho] - {\cal D}^{\rm
CM}[\varrho] \|}{ \frac{n m d^4 U_0^2 \|F\|^2}{\hbar^2 \sqrt{mkT}}
} \sim \frac{1}{kT}\max\left(|U_0|\|F\|, \frac{\hbar^2}{md^2} \right).
\end{equation}

It is also possible to slightly adapt the CM approach to allow for
lowering velocity of gas particles by considering a perturbation
of their trajectories from straight lines caused by a
state-dependent potential $\ave{F} U({\bf r})$, where $\ave{F} =
{\rm tr}[F \sum_i \mu_i \ket{i}\bra{i} \otimes \varrho_S] = \sum_i
\mu_i {\rm tr}[A_{ii} \varrho_S]$. For a spherical square-well
potential with negative $\ave{F} U_0$ the perturbed trajectory
consists of 3 line segments. The angle of incidence $\alpha$ and
the angle of refraction $\beta$ at the first vertex satisfy the
relation $p \sin\alpha = p' \sin\beta$, where $p$ and
$p'=\sqrt{p^2 + 2m \ave{F} |U_0|}$ are the momenta of the particle
outside and inside of the region $|{\bf r}| \leq d$, respectively.
Additionally, the angle of incidence is related to the impact
parameter $b$ by formula $\sin\alpha = \frac{b}{d}$. The effective
collision time
\begin{equation} \label{tau-corrected}
\tau = \frac{2md\cos\beta}{p'} = \frac{2m\sqrt{ (d^2-b^2)p^2 + 2m
\ave{F} |U_0|d^2 }}{p^2 + 2m \ave{F} |U_0|}.
\end{equation}

\noindent Using the exact expression~\eqref{tau-corrected} for
$\tau$, we find the coefficients $\left\langle \frac{U_0
\tau}{t_{\rm free}} \right\rangle$ and $\left\langle \frac{U_0^2
\tau^2}{\hbar^2 t_{\rm free}} \right\rangle$ in the Lamb
shift~\eqref{Lamb-CM} and the dissipator~\eqref{D-CM} in the
modified semiclassical collision model. The first order expansion
of these coefficients with respect to small parameter
$\frac{|U_0|}{kT}$ reads
\begin{eqnarray}
&& \!\!\!\!\!\!\!\!\!\! \label{C1-rectangular-corrected}
\left\langle \frac{U_0 \tau}{t_{\rm free}}
\right\rangle = \frac{4\pi}{3} nd^3 U_0 \left[ 1 - \frac{\ave{F} U_0}{kT} + o\left(\frac{\ave{F}|U_0|}{kT}\right) \right],\\
&& \!\!\!\!\!\!\!\!\!\! \label{C2-rectangular-corrected}
\left\langle \frac{U_0^2 \tau^2}{\hbar^2 t_{\rm free}}
\right\rangle = \frac{2\sqrt{2\pi} n m d^4 U_0^2}{\hbar^2
\sqrt{mkT}} \left[ 1 + \frac{\ave{F} U_0}{kT} +
o\left(\frac{\ave{F}|U_0|}{kT}\right) \right]. \nonumber\\
\end{eqnarray}

\noindent We see that such a non-linear modification of the
collision model provides a better agreement between
$\widetilde{H}_{\rm LS}^{\rm LDL}$ and $H_{\rm LS}^{\rm CM}$ and
between $\widetilde{\cal D}^{\rm LDL}$ and ${\cal D}^{\rm CM}$ for
some states $\varrho_S$ and operators $F$, however, the
discrepancy between the two approaches is still given by
formulas~\eqref{H-H} and~\eqref{D-D} in general.

\section{Conclusions} \label{section-conclusions}

We developed and compared two approaches to the analysis of the
open quantum system dynamics induced by interaction of the
spin-like system with a dilute gas of spin-like particles with
internal degrees of freedom: the low density limit in the fully
quantum scenario and the semiclassical collision model. We derived
GKSL master equations for a specific class of system-particle
interaction Hamiltonians of the form $\widetilde{H}_{S1} = F
\otimes U({\bf r})$, however, the results remain valid for a
general spin-dependent scattering process with the interaction
Hamiltonian $\widetilde{H}_{S1} = \sum_{i,j,k,l} \ket{k}\bra{l}
\otimes \ket{i}\bra{j} \otimes F_{ki,lj}({\bf r})$. Using the
first-order Born approximation in the fully quantum treatment, the
simplified expressions for the Lamb shift~\eqref{Lamb-LDL} and the
dissipator~\eqref{D-LDL} were derived. In the semiclassical
collision model, we used the approximation of straight
trajectories and the stroboscopic approximation to get the Lamb
shift~\eqref{Lamb-CM} and the dissipator~\eqref{D-CM}. We proved
equivalence of the Lamb shifts in both approaches and found that
both dissipators~\eqref{D-LDL} and \eqref{D-CM} qualitatively
coincide for finite temperatures and quantitatively coincide in
the limit $T \rightarrow \infty$. The illustrative examples of
Gaussian and spherical square-well potentials are considered, for
which the dissipators~\eqref{D-LDL} and \eqref{D-CM} are compared
in the case of fast particles up to the second order of the
scattering potential $F \otimes U({\bf r})$. The sufficient
conditions for the two approaches to give the same master equation
are $n d^3\ll 1$, $kT \gg \frac{\hbar^2}{md^2}$, and $U_0 \ll
\sqrt{\frac{\hbar^2 kT}{m d^2}}$.

\begin{acknowledgements}
The study in Secs. II.B, II.C, III.B, IV, and V was supported by
the Russian Science Foundation under Project No. 17-11-01388 and
performed in Steklov Mathematical Institute of Russian Academy of
Sciences. Secs. I, II.A, and III.A were written in Valiev
Institute of Physics and Technology of Russian Academy of
Sciences, where S.N.F. was supported by Program No. 0066-2019-0005
of the Russian Ministry of Science and Higher Education; Moscow
Institute of Physics and Technology, where S.N.F. and G.N.S. were
supported by the Foundation for the Advancement of Theoretical
Physics and Mathematics ``BASIS'' under Grant No. 19-1-2-66-1; and
National University of Science and Technology ``MISIS'', where
A.N.P. was supported by Project No. 1.669.2016/1.4 of the Russian
Ministry of Science and Higher Education.
\end{acknowledgements}


\begin{thebibliography}{56}
\bibitem{breuer-2002}
H.-P. Breuer and F. Petruccione, \emph{The Theory of Open Quantum
Systems} (Oxford University Press, Oxford, 2002).

\bibitem{schoeller-2018}
H. Schoeller, Dynamics of open quantum systems, arXiv:1802.10014
(2018).

\bibitem{cui-2006}
P. Cui, X.-Q. Li, J. Shao, and Y. Yan, Quantum transport from the
perspective of quantum open systems, Phys. Lett. A {\bf 357}, 449
(2006).

\bibitem{talarico-2019}
N. W. Talarico, S. Maniscalco, and N. Lo Gullo, A scalable
numerical approach to the solution of the Dyson equation for the
non-equilibrium single-particle Green's function, Phys. Status
Solidi B {\bf 256}, 1800501 (2019).

\bibitem{valkunas-2013}
L.~Valkunas, D.~Abramavicius, and T.~Mancal, \emph{Molecular
Excitation Dynamics and Relaxation: Quantum Theory and
Spectroscopy} (Wiley, New York, 2013).

\bibitem{degen-2017}
C.~L. Degen, F.~Reinhard, and P.~Cappellaro, Quantum sensing, Rev.
Mod. Phys. {\bf 89}, 035002 (2017).

\bibitem{wilde-2017}
M.~M. Wilde, \emph{Quantum Information Theory} (Cambridge
University Press, Cambridge, England, 2017).

\bibitem{nielsen-2000}
M.~A. Nielsen and I.~L. Chuang, \emph{Quantum Computation and
Quantum Information} (Cambridge University Press, Cambridge,
England, 2000).

\bibitem{filippov-2019}
S. N. Filippov, Quantum mappings and characterization of entangled
quantum states, J. Math. Sci. {\bf 241}, 210 (2019).

\bibitem{glaser-2015}
S. J. Glaser, U. Boscain, T. Calarco, C. P. Koch, W.
K\"{o}ckenberger, R. Kosloff, I. Kuprov, B. Luy, S. Schirmer, T.
Schulte-Herbr\"{u}ggen, D. Sugny, and F. K. Wilhelm, Training
Schr\"{o}dinger's cat: quantum optimal control, Eur. Phys. J. D
69, 279 (2015).

\bibitem{rivas-2014}
\'{A}. Rivas, S. F. Huelga, and M. B. Plenio, Quantum
non-Markovianity: characterization, quantification and detection,
Rep. Prog. Phys. {\bf 77}, 094001 (2014).

\bibitem{breuer-2016}
H.-P. Breuer, E.-M. Laine, J. Piilo, and B. Vacchini, Colloquium:
Non-Markovian dynamics in open quantum systems, Rev. Mod. Phys.
{\bf 88}, 021002 (2016).

\bibitem{de-vega-2017}
I. de Vega and D. Alonso, Dynamics of non-Markovian open quantum
systems, Rev. Mod. Phys. {\bf 89}, 015001 (2017).

\bibitem{benatti-2017}
F. Benatti, D. Chru\'{s}ci\'{n}ski, and S. Filippov, Tensor power
of dynamical maps and positive versus completely positive
divisibility, Phys. Rev. A {\bf 95}, 012112 (2017).

\bibitem{fc-2018}
S. N. Filippov and D. Chru\'{s}ci\'{n}ski, Time deformations of
master equations, Phys. Rev. A {\bf 98}, 022123 (2018).

\bibitem{li-2018}
L.~Li, M.~J.~W. Hall, and H.~M. Wiseman, Concepts of quantum
non-Markovianity: A hierarchy, Phys. Rep. {\bf 759}, 1 (2018).

\bibitem{luchnikov-2019}
I. A. Luchnikov, S. V. Vintskevich, H. Ouerdane, and S.~N.
Filippov, Simulation complexity of open quantum dynamics:
Connection with tensor networks, Phys. Rev. Lett. {\bf 122},
160401 (2019).

\bibitem{van-hove-1954}
L. van Hove, Quantum-mechanical perturbations giving rise to a
statistical transport equation, Physica {\bf 21}, 517 (1954).

\bibitem{davies-1974}
E. B. Davies, Markovian master equations, Commun. Math. Phys. {\bf
39}, 91 (1974).

\bibitem{spohn-1978}
H. Spohn and J. L. Lebowitz, Irreversible thermodynamics for
quantum systems weakly coupled to thermal reservoirs, Adv. Chem.
Phys. {\bf 38}, 109 (1978).

\bibitem{accardi-1990}
L. Accardi, A. Frigerio, and Y. G. Lu, The weak coupling limit as
a quantum functional central limit, Comm. Math. Phys. {\bf 131},
537 (1990).

\bibitem{palmer-1977}
P. F. Palmer, The singular coupling and weak coupling limits, J.
Math. Phys. {\bf 18}, 527 (1977).

\bibitem{gorini-1978}
V. Gorini, A. Frigerio, M. Verri, A. Kossakowski, and E. C. G.
Sudarshan, Properties of quantum Markovian master equations, Rep.
Math. Phys. {\bf 13}, 149 (1978).

\bibitem{accardi-book}
L. Accardi, Y. G. Lu, and I. Volovich, \emph{Quantum Theory and
Its Stochastic Limit} (Springer, Berlin, 2002).

\bibitem{pechen-2002}
A. N. Pechen and I. V. Volovich, Quantum multipole noise and
generalized quantum stochastic equations, Infinite Dimensional
Analysis, Quantum Probability and Related Topics {\bf 5}, 441
(2002).

\bibitem{dumcke-1985}
R. D\"{u}mcke, The low density limit for an $N$-level system
interacting with a free Bose or Fermi gas, Commun. Math. Phys.
{\bf 97}, 331 (1985).

\bibitem{accardi-1991}
L. Accardi and Y. G. Lu, The low-density limit of quantum systems,
J. Phys. A: Math. Gen. {\bf 24}, 3483 (1991).

\bibitem{accardi-1992}
L. Accardi and Y. Lu, The low density limit in finite temperature
case, Nagoya Mathematical Journal {\bf 126}, 25 (1992).

\bibitem{rudnicki-1992}
S. Rudnicki, R. Alicki, and S. Sadowski, The low-density limit in
terms of collective squeezed vectors, J. Math. Phys. {\bf 33},
2607 (1992).

\bibitem{apv-2002}
L Accardi, A. N. Pechen, and I. V. Volovich, Quantum stochastic
equation for the low density limit, J. Phys. A: Math. Gen. {\bf
35}, 4889 (2002).

\bibitem{accardi-2003}
L. Accardi, A. N. Pechen, and I. V. Volovich, A stochastic golden
rule and quantum Langevin equation for the low density limit,
Infinite Dimensional Analysis, Quantum Probability and Related
Topics {\bf 6}, 431 (2003).

\bibitem{pechen-2004}
A. N. Pechen, Quantum stochastic equation for a test particle
interacting with a dilute Bose gas, J. Math. Phys. {\bf 45}, 400
(2004).

\bibitem{pechen-jmp-2006}
A. N. Pechen, The multitime correlation functions, free white
noise, and the generalized Poisson statistics in the low density
limit, J. Math. Phys. {\bf 47}, 033507 (2006).

\bibitem{rau-1963}
J. Rau, Relaxation phenomena in spin and harmonic oscillator
systems, Phys. Rev. {\bf 129}, 1880 (1963).

\bibitem{giovannetti-2012}
V. Giovannetti and G. M. Palma, Master Equations for Correlated
Quantum Channels, Phys. Rev. Lett. {\bf 108}, 040401 (2012).

\bibitem{lorenzo-2017}
S. Lorenzo, F. Ciccarello, and G. M. Palma, Composite quantum
collision models, Phys. Rev. A {\bf 96}, 032107 (2017).

\bibitem{luchnikov-2017}
I. A. Luchnikov and S. N. Filippov, Quantum evolution in the
stroboscopic limit of repeated measurements, Phys. Rev. A {\bf
95}, 022113 (2017).

\bibitem{hornberger-2007}
K. Hornberger, Monitoring approach to open quantum dynamics using
scattering theory, EPL {\bf 77}, 50007 (2007).

\bibitem{hornberger-2008}
K. Hornberger and B. Vacchini, Monitoring derivation of the
quantum linear Boltzmann equation, Phys. Rev. A {\bf 77}, 022112
(2008).

\bibitem{vacchini-2009}
B. Vacchini and K. Hornberger, Quantum linear Boltzmann equation,
Phys. Rep. {\bf 478}, 71 (2009).

\bibitem{smirne-2010}
A. Smirne and B. Vacchini, Quantum master equation for collisional
dynamics of massive particles with internal degrees of freedom,
Phys. Rev. A {\bf 82}, 042111 (2010).

\bibitem{gks-1976}
V. Gorini, A. Kossakowski, and E.~C.~G. Sudarshan, Completely
positive dynamical semigroups of $N$-level systems, J. Math. Phys.
{\bf 17}, 821 (1976).

\bibitem{lindblad-1976}
G.~Lindblad, On the generators of quantum dynamical semigroups,
Comm. Math. Phys. {\bf 48}, 119 (1976).

\bibitem{martinetz-2018}
L. Martinetz, K. Hornberger, and B. A. Stickler, Gas-induced
friction and diffusion of rigid rotors, Phys. Rev. E {\bf 97},
052112 (2018).

\bibitem{wineland-1998}
D. J. Wineland, C. Monroe, W. M. Itano, D. Leibfried, B. E. King,
and D. M. Meekhof, Experimental issues in coherent quantum-state
manipulation of trapped atomic ions, J. Res. Natl. Inst. Stand.
Tech. {\bf 103}, 259 (1998).

\bibitem{serra-2001}
R. M. Serra, N. G. de Almeida, W. B. da Costa, and M. H. Y.
Moussa, Decoherence in trapped ions due to polarization of the
residual background gas, Phys. Rev. A {\bf 64}, 033419 (2001).

\bibitem{uys-2005}
H. Uys, J. D. Perreault, and A. D. Cronin, Matter-wave decoherence
due to a gas environment in an atom interferometer, Phys. Rev.
Lett. {\bf 95}, 150403 (2005).

\bibitem{pechen-2006}
A. Pechen and H. Rabitz, Teaching the environment to control
quantum systems, Phys. Rev. A {\bf 73}, 062102 (2006).

\bibitem{pechen-rabitz-2014}
A. Pechen and H. Rabitz, Incoherent control of open quantum
systems, J. Math. Sci. {\bf 199}, 695 (2014).

\bibitem{pechen-2019}
A. N. Pechen', Some mathematical problems of control of quantum
systems, J. Math. Sci. {\bf 241}, 185 (2019).

\bibitem{alicki-2003}
R. Alicki and S. Kryszewski, Completely positive Bloch-Boltzmann
equations, Phys. Rev. A {\bf 68}, 013809 (2003).

\bibitem{koniorczyk-2008}
M. Koniorczyk, \'{A}. Varga, P. Rap\v{c}an, and V. Bu\v{z}ek,
Quantum homogenization and state randomization in semiquantal spin
systems, Phys. Rev. A {\bf 77}, 052106 (2008).

\bibitem{rempe-1990}
G. Rempe, F. Schmidt-Kaler, and H. Walther, Observation of
sub-Poissonian photon statistics in a micromaser, Phys. Rev. Lett.
{\bf 64}, 2783 (1990).

\bibitem{scarani-2002}
V. Scarani, M. Ziman, P. \v{S}telmachovi\v{c}, N. Gisin, and V.
Bu\v{z}ek, Thermalizing quantum machines: Dissipation and
entanglement, Phys. Rev. Lett. {\bf 88}, 097905 (2002).

\bibitem{rybar-2012}
T. Ryb\'{a}r, S. N. Filippov, M. Ziman, and V. Bu\v{z}ek,
Simulation of indivisible qubit channels in collision models, J.
Phys. B: At. Mol. Opt. Phys. {\bf 45}, 154006 (2012).

\bibitem{mccloskey-2014}
R. McCloskey and M. Paternostro, Non-Markovianity and
system-environment correlations in a microscopic collision model,
Phys. Rev. A {\bf 89}, 052120 (2014).

\bibitem{kretschmer-2016}
S. Kretschmer, K. Luoma, and W. T. Strunz, Collision model for
non-Markovian quantum dynamics, Phys. Rev. A {\bf 94}, 012106
(2016).

\bibitem{dabrowska-2017}
A. D\k{a}browska, G. Sarbicki, and D. Chru\'{s}ci\'{n}ski, Quantum
trajectories for a system interacting with environment in a
single-photon state: Counting and diffusive processes, Phys. Rev.
A {\bf 96}, 053819 (2017).

\bibitem{filippov-2017}
S. N. Filippov, J. Piilo, S. Maniscalco, and M. Ziman,
Divisibility of quantum dynamical maps and collision models, Phys.
Rev. A {\bf 96}, 032111 (2017).

\bibitem{ciccarello-2017}
F. Ciccarello, Collision models in quantum optics, Quantum
Measurements and Quantum Metrology {\bf 4}, 53 (2017).

\bibitem{levy-2012}
A. Levy, R. Alicki, and R. Kosloff, Quantum refrigerators and the
third law of thermodynamics, Phys. Rev. E {\bf 85}, 061126 (2012).

\bibitem{kosloff-2013}
R. Kosloff, Quantum thermodynamics: A dynamical viewpoint, Entropy
{\bf 15}, 2100 (2013).

\bibitem{kosloff-2019}
R. Kosloff, Quantum thermodynamics and open-systems modeling, J.
Chem. Phys. {\bf 150}, 204105 (2019).

\bibitem{LL}
L. D. Landau and E. M. Lifshitz, \emph{Quantum Mechanics:
Non-Relativistic Theory}, Sec. 125 (Pergamon, London, 1965).

\end{thebibliography}
\end{document}